\documentclass[a4paper]{jpconf}
\usepackage{graphicx}
\usepackage{amsmath}
\begin{document}
\title{
Cluster dynamical mean-field study of the Hubbard model 
on a 3D frustrated hyperkagome lattice}

\author{Masafumi Udagawa and Yukitoshi Motome}

\address{Department of Applied Physics, University of Tokyo,
7-3-1 Hongo, Bunkyo-ku, Tokyo, Japan}

\ead{udagawa@ap.t.u-tokyo.ac.jp}

\begin{abstract}
We study the Hubbard model on a geometrically-frustrated hyperkagome lattice 
 by a cluster extension of the dynamical mean field theory. We calculate the
 temperature ($T$) dependences of the specific heat ($C$) and the spin-lattice
 relaxation time ($T_1$) in correlated metallic region. 
 $C/T$ shows a peak at $T=T_{{\text p}1}$ and 
 rapidly decreases as $T\rightarrow 0$. 
 On the other hand, $1/T_1T$ has a peak at a higher
 temperature $T_{{\text p}2}$ than $T_{{\text p}1}$, 
 and largely decreases below
 $T_{{\text p}2}$, followed by the Korringa law $1/T_1 \propto T$  as $T\rightarrow 0$. Both peak temperatures are suppressed and the peaks become sharper
 as electron correlation is increased.
 These behaviors originate from 
 strong renormalization of the energy scales
 in the peculiar electronic structure in this frustrated system;
 a pseudo-gap like feature, the van-Hove singularity, and the flat band.
 The results are discussed in comparison with 
 the experimental data in the hyperkagome material, Na$_4$Ir$_3$O$_8$.
 \end{abstract}

\section{Introduction}
\label{sec:intro}
How the electronic structure and thermodynamic properties evolve
through the Mott transition has been a central issue in condensed matter physics
\cite{Imada1998}.
Recently, insulating states close to the Mott transition have attracted
considerable attention due to their exotic properties.
Typical examples are found in a 2D organic conductor \cite{Shimizu03}
and $^3$He adsorbed on graphite \cite{Ishida97}:
Both are considered to be in the vicinity of the Mott transition
having a tiny charge gap, and 
exhibit gapless spin-liquid behavior at low temperatures.
If electron interaction is sufficiently strong and 
there is a large charge gap, a localized spin model gives 
a good description of the systems.
Whereas, in the vicinity of the Mott transition, 
the expansion from the strong coupling limit 
suffers from slow convergence due to the small charge gap. 
In order to describe such systems, it is crucial to take account of charge fluctuations
as well as spin ones.

Among such insulators near the Mott transition showing exotic behaviors, 
we focus on Na$_4$Ir$_3$O$_8$ in this paper. 
The lattice structure of Na$_4$Ir$_3$O$_8$ is similar to that of spinel oxides,
while $1/4$ of Ir$^{4+}$ pyrochlore sites are regularly 
replaced by non-magnetic Na$^+$ cations \cite{Okamoto07}. 
The Ir$^{4+}$ sublattice consists of a highly-frustrated three-dimensional (3D) 
network of corner-sharing triangles, which is called the `hyperkagome' lattice
[Fig.~\ref{fig:DOS_U0}(a)]. 
While the hyperkagome lattice shares a common local connectivity to 
the 2D kagome lattice, it has several distinct features 
such as the three dimensionality and structural chirality. 
It was reported that the charge gap of Na$_4$Ir$_3$O$_8$ is fairly small,
suggesting that the compound is near the Mott transition: 
The magnitude of the gap is estimated at $\sim 500$ K \cite{Okamoto07}.

Puzzling behaviors have been reported for the
thermodynamic properties of Na$_4$Ir$_3$O$_8$ \cite{Okamoto07}. 
The Curie-Weiss temperature is large $\sim 650$ K,
which is comparable to the magnitude of the charge gap.
In spite of the large Curie-Weiss temperature,
there is no clear sign of phase transition down to $2$ K. 
$C/T$ shows a peak at $T\sim 20$ K, and rapidly decreases as $T\rightarrow 0$.
In contrast, the magnetic susceptibility takes a finite value as $T\rightarrow 0$, 
implying a macroscopic number of gapless magnetic excitations. 
As to the spin dynamics, the spin-lattice relaxation time, $T_1$, shows 
anomalous temperature dependence: $1/T_1T$ shows a small peak at $T\sim 250$ K, and
decreases down to $T\sim 50$ K. Below $T\sim 50$ K, $1/T_1T$
shows a moderate temperature dependence, seemingly obeying the Korringa law \cite{Fujiyama_private}.

In spite of several theoretical efforts to make 
sense of the wide breadth of experimental results,
comprehensive understanding has not been reached yet. 
Instabilities toward spin nematic phase and $Z_2$ spin liquid were discussed 
for the classical and quantum Heisenberg model on
the hyperkagome lattice \cite{Hopkinson07, Lawler07, Zhou08}. 
Effects of the Dzyaloshinskii-Moriya interaction and 
anisotropic exchange interaction
have also been studied \cite{Gang08}. 

A part of the difficulties faced by the theoretical studies stems from
the fact that this compound locates in the vicinity of the Mott transition. 
The exchange interaction estimated from the Curie-Weiss temperature
is comparable to the magnitude of the charge gap, 
which suggests an importance of including charge fluctuations in addition to spin ones.
So far, however, all the theoretical analyses have been done for localized spin models.
In this paper, in order to gain a new insight into this puzzling situation,
we consider the problem from the opposite side, i.e., from the metallic region.
We will explore how the electronic structure evolves 
as the electron correlation increases in the 3D frustrated hyperkagome systems.

\section{Model and Method}
We study the Hubbard model on the 3D frustrated hyperkagome lattice
shown in Fig.~\ref{fig:DOS_U0}(a),
whose Hamiltonian is given by
\begin{eqnarray}
\mathcal{H} = -t \sum\limits_{\langle i,j\rangle,
 \sigma} \bigl(c^{\dagger}_{i\sigma}c_{j\sigma} + {\text{h.c.}}\bigr) +
 U\sum\limits_i n_{i\uparrow}n_{i\downarrow} - \mu\sum\limits_{i\sigma}n_{i\sigma},
\end{eqnarray}
in the standard notations.
We consider only the nearest-neighbor hopping and set $t=1$ as an energy unit hereafter.
The chemical potential $\mu$ is controlled so that the system is at the half filling
(one electron per site on average).
In Na$_4$Ir$_3$O$_8$, each Ir$^{4+}$ cation has 
one electron in three-fold $t_{2g}$ orbitals on average. 
Our model may describe
the case that the trigonal distortion splits the $t_{2g}$ levels and 
the lower nondegenerate $a_{1g}$ level plays a major role.

To investigate the thermodynamics of this model, 
we adopt a cluster extension of the dynamical mean field theory \cite{Kotliar2001}. 
In this method, the original lattice problem is
mapped to a cluster impurity problem. 
In the present case, in order to take account of the unique spatial geometry of 
the hyperkagome structure shown in Fig.~\ref{fig:DOS_U0}(a),
we consider a 12-sites cluster (the unit cell of the hyperkagome lattice), and
apply the iterative perturbation theory (IPT) as an impurity solver. 
IPT is known to give a reliable description including 
the Mott transition for a single-impurity problem \cite{Georges92},
while it becomes poor for large $U$ in the case of clusters.
In the present study, we limit our investigations
in the small to intermediate $U$ regime 
where the method is more reliable.

\section{Results}
\begin{figure}[h]
\begin{minipage}[t]{20pc}
\includegraphics[width=18pc]{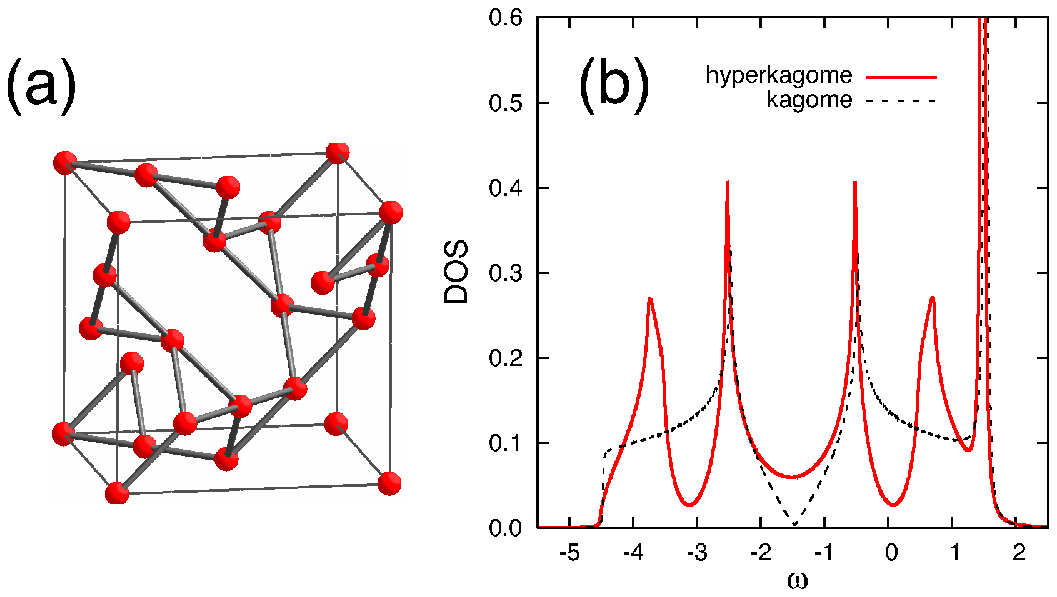}
\caption{\label{fig:DOS_U0}
(a) Unit cell of the hyperkagome lattice.
(b) Density of states of the noninteracting tight-binding model
on the 3D hyperkagome lattice in comparison with that on the 2D kagome lattice.
Fermi energy is set at $\omega=0$.}
\end{minipage}\hspace{2pc}%
\begin{minipage}[t]{14pc}
\includegraphics[width=10.46pc]{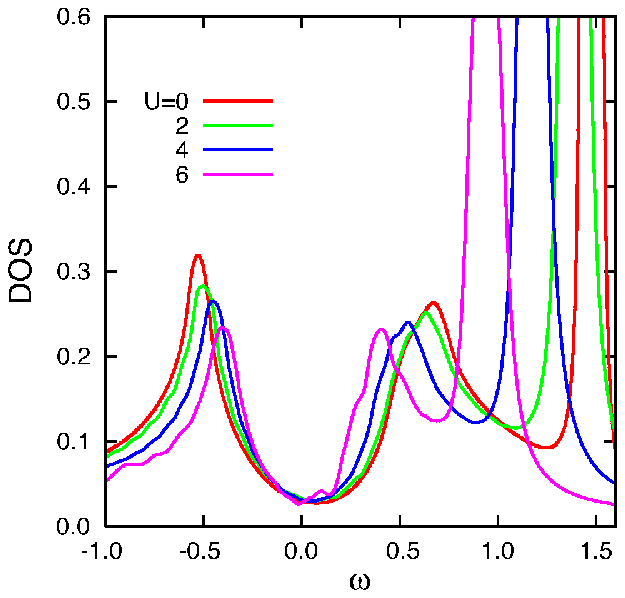}
\caption{\label{fig:DOS_T0.1}
$U$ dependence of the density of states near the Fermi energy at $T=0.1$.
}
\end{minipage} 
\end{figure}

First let us consider the basic electronic structure of the hyperkagome model.
In Fig.~\ref{fig:DOS_U0}(b), we show the density of states (DOS) at $U=0$
in comparison with that of the 2D kagome case.
The diverging peak at the upper band edge ($E_{\text{flat}} \simeq 1.52$) 
comes from four-fold--degenerate flat bands. 
In addition, two sharp peaks exist in the middle of the band, which are the van-Hove singularities at
$E_{\text{v}1} \simeq -0.48$ and $E_{\text{v}2} \simeq -2.48$, respectively. 
These features are similar to the 2D kagome case, 
however, there is an important difference:
DOS of the hyperkagome model has a large dip near the Fermi energy ($\omega = 0$), 
showing a pseudo-gap like feature. 
As a consequence, DOS at the Fermi energy is largely suppressed
down to $\sim 20$\% of that in the 2D kagome case.
This small DOS might be relevant to consider the fact
that Na$_4$Ir$_3$O$_8$ is an insulator even though the electron correlation is expected 
to be weak for $5d$ electrons of Ir$^{4+}$ cations.

When we switch on the electron correlation $U$, these features of DOS 
near the Fermi energy undergo a strong renormalization by $U$.
We show the results at low $T$ in Fig.~\ref{fig:DOS_T0.1}.
Energies of the flat band and the van-Hove singularity, $E_{\text{flat}}$ and
$E_{\text{v1}}$, are renormalized and approach the Fermi energy.
These energy scales and their renormalization give rise to
peculiar temperature dependences as we will see below.

\begin{figure}[h]
\begin{minipage}[t]{17pc}
\includegraphics[width=13.2pc]{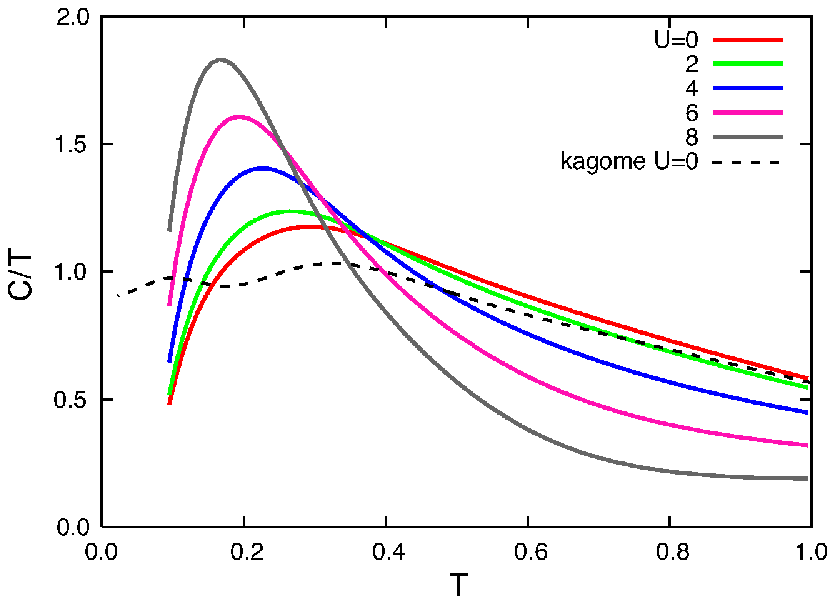}
\caption{\label{CoverT}Temperature dependence of $C/T$. 
The dashed line is the result at $U=0$ for the 2D kagome case for comparison.}
\end{minipage}\hspace{2pc}%
\begin{minipage}[t]{16pc}
\includegraphics[width=13pc]{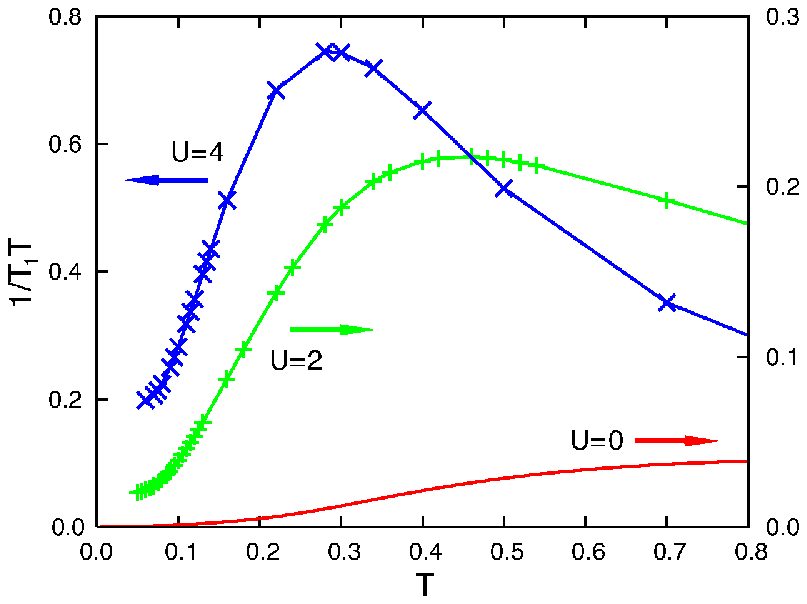}
\caption{\label{T1T}Temperature dependence of $1/T_1T$.}
\end{minipage} 
\end{figure}

Next, we show the results for the specific heat ($C$) in Fig.~\ref{CoverT}.
At $U=0$, $C/T$ has a peak at $T = T_{{\text p}1} \sim 0.3$, and 
rapidly decreases for $T < T_{{\text p}1}$.
$T_{\text{p}1}$ correlates with the energy scale of the van-Hove singularity
closer to the Fermi level, $E_{\text{v}1}$: 
When $T$ crosses $|E_{\text{v}1}|$, $C$ shows a peak. 
The rapid decrease at lower $T$ is ascribed to
the small DOS at the Fermi level. 
For comparison, we show $C/T$ for the 2D kagome model at $U=0$ in Fig.~\ref{CoverT},
which does not show such rapid decrease.
As $U$ increases, $T_{{\text p}1}$ shifts to lower $T$ and 
the peak becomes sharper. 
As a result, the entropy is released more rapidly 
at low $T$ for larger $U$.
These interesting behaviors are closely related to 
the renormalization of DOS by $U$ shown in Fig.~\ref{fig:DOS_T0.1}. 
We note that the behaviors of $C/T$ are similar to those observed in the experiment in Na$_4$Ir$_3$O$_8$ \cite{Okamoto07}. 

Finally, we discuss the spin relaxation time $T_1$. 
We calculate $1/T_1T$ from the standard
formula, $1/T_1T \propto \lim_{\omega\to0} \sum_{\mathbf q} \text{Tr} \ \text{Im}\hat{\chi}^{+-}({\mathbf q}, \omega + i0) / \omega N$. 
In order to obtain the transverse spin susceptibility, $\hat{\chi}^{+-}({\mathbf q}, \omega +
i0)$, we approximate the irreducible vertex function 
in the Bethe-Salpeter equation by the RPA contribution.
Figure~\ref{T1T} shows the results. 
$1/T_1T$ shows a broad peak and decreases as $T$ is lowered. 
At low $T$ below $\sim 0.05$, it approaches a constant value, obeying the Korringa law. 
Although the peak shifts to lower $T$ and the peak becomes sharper
for larger $U$ as seen in $C/T$, the peak temperature, $T_{{\text p}2}$, is different from $T_{{\text p}1}$ and 
higher than $T_{{\text p}1}$. 
This might be attributed to the fact that
$1/T_1T$ is more sensitive to the flat band at a higher energy
whose singularity is much stronger than the van-Hove singularity.
Note that 
$1/T_1T$ is proportional to the square of DOS at $U=0$, 
in contrast to $C/T$ which is proportional to DOS itself.
Our results show similar features to $1/T_1T$ of Na$_4$Ir$_3$O$_8$ 
such as a peak at high $T$ followed by a large drop and a Korringa-like behavior at the lowest $T$ \cite{Fujiyama_private}.

\section{Discussion and Concluding Remarks}
We have obtained characteristic temperature dependences of $C/T$ and $1/T_1T$:
Both show a peak followed by a rapid decrease at low $T$, and
the peaks shift to lower $T$ as increasing $U$.
These behaviors are explained by the strong renormalization effect by $U$ on
the peculiar structure of DOS in the hyperkagome model.
Particularly important factors are the pseudo-gap like feature and
two singularities, the van-Hove singularity and the flat band.

It is rather surprising that these results in the metallic regime 
are qualitatively similar to those observed in the hyperkagome material, 
Na$_4$Ir$_3$O$_8$, which is an insulator with a small charge gap.
The agreement implies a possibility that
the renormalized singularities of DOS survive and
affect the thermodynamic properties even in the insulating regime 
when the system is in the vicinity of the Mott transition.
Note that although the Mott transition itself is known to be of first order,
it is like a liquid-gas transition where the finite-temperature properties
can be smoothly connected by bypassing the critical endpoint \cite{Imada1998}.

To explore this fascinating scenario,
it is necessary to 
extend our analysis to the insulating region beyond the Mott transition. Such study by employing an improved impurity solver is in progress.
There, it is interesting to examine 
whether the vanishing or very tiny $C/T$,
the Korringa-like behavior of $1/T_1T$, and
a finite value of the magnetic susceptibility at $T\to 0$ 
are comprehensively explained.
Considering the real compound,
it is also interesting to consider effects of orbital degrees of freedom, 
spin-orbit interaction, and disorder due to Na deficiency. 

We thank S. Fujiyama and H. Takagi for fruitful discussions
and for showing their NMR data prior to publication.
This work was supported by Grant-in-Aid for Scientific Research on Priority Areas 
`Physics of New Quantum Phases in SuperClean Materials' (17071003) and
`Novel States of Matter Induced by Frustration' (19052008).

\end{document}